\documentclass[10pt,prl,showpacs,superscriptaddress,twocolumn,citeautoscript]{revtex4-1}

\usepackage{graphicx,color}  
\usepackage{dcolumn}   
\usepackage{bm}        
\usepackage{amssymb,url}
\usepackage{amsmath}
\usepackage{hyperref}
\usepackage{braket}

\newcommand{\beq}{\begin{equation}}
\newcommand{\eeq}{\end{equation}}

\begin{document}

\title{Engineered diffraction gratings for acoustic cloaking}
\author{Yabin Jin}
\affiliation{School of Aerospace Engineering and Applied Mechanics and Institute for Advanced Study, Tongji University, 200092 Shanghai, People's Republic of China}
\author{Xinsheng Fang}
\author{Yong  Li}
\email{yongli@tongji.edu.cn }
\affiliation{Institute of Acoustics, School of Physics Science and Engineering, Tongji University, Shanghai 200092, People's Republic of China}
\affiliation{Shanghai Key Laboratory of Special Artificial Microstructure Materials and Technology, School of Physics Science and Engineering, Tongji University, Shanghai 200092, People's Republic of China}
\author{Daniel Torrent}
\email{dtorrent@uji.es}
\affiliation{GROC, UJI, Institut de Noves Tecnologies de la Imatge (INIT), Universitat Jaume I, 12071, Castell\'o, (Spain)}
\date{\today}

\begin{abstract}
We show that engineered diffraction gratings can considerably simplify the design and performance of acoustic devices. Acoustic reflecting gratings are designed in such a way that all the incident energy is channeled towards the diffracted mode traveling in the oposite direction of the incident field (retroreflection effect), and this effect is used to cloak an object placed over an acoustically rigid surface. Axisymmetric gratings consisting in rigid surfaces with just one groove per unit cell are used to design thin acoustic carpet cloaks. Finally, full wave numerical simulations are performed and a conical carpet cloak is experimentally tested, showing an excellent scattering cancellation effect.
 \end{abstract}

\maketitle
Acoustic cloaking\cite{cummer2007one} is one of the most challenging problems related with classical wave control. Directly related with the original idea of electromagnetic cloaking\cite{pendry2006controlling}, a huge amount of research works have been devoted to theoretically and experimentally solve this problem, being the requirement of singular parameters one of the major milestones to overcome the realization of these devices. The so-called carpet cloak\cite{li2008hiding}, or ground cloak, is a special type of cloaking shell making objects invisible positioned on a flat mirror, with the remarkable property of not requiring extreme materials' parameters for their realization. Acoustic carpet cloaks with finite anisotropy and homogeneous materials' parameters can be designed in the framework of acoustic metamaterials. For instance, perforated plastic plates\cite{popa2011experimental, zigoneanu2014three} and steel/air composites\cite{zhang2012feasible} have been proposed to numerically and experimentally study 2D and 3D carpet cloaks for air-borne acoustic wave propagation. In underwater acoustics, layered mercury inclusions\cite{xiong2015design} or brass plates\cite{bi2017design} were considered to theoretically design carpet cloaks, which have been recently experimentally demonstrated by designing a steel stripe-composited pyramid\cite{bi2018experimental}. Finally, carpet cloaks have also been explored for other type of mechanical waves, like water \cite{wang2017carpet} or elastic\cite{zhang2018hyperelastic} waves. However, the size of carpet cloak shell compared to that of the cloaked region is still bulky, what limits the potential applications of these devices.

With the advent of metasurfaces \cite{yu2011light,xie2014wavefront,li2013reflected}, which are artificially structured thin surfaces capable of modulate the reflected wavefronts, a new type of carpet cloaks were envisioned. Then, thin carpet cloaks based on Helmholtz resonators\cite{yang2016metasurface,faure2016experiments,dubois2017thin}, membranes \cite{esfahlani2016acoustic, zhai2016ultrathin} and spiral cavities\cite{wang2017broadband} were designed in this framework. However, the main constraint of these structures is that metasurfaces are complex periodically structured surfaces in which the complexity of the unit cell strongly hinders their effectivity, since only very small cloaks with a small number of periods are feasible from the practical point of view. Additionally, the need of a large number of resonators enforces the device to work in the deep sub-wavelength and dissipation use to be a major drawback in their functionality.


In this work an engineered grating\cite{torrent2018acoustic} will be applied to the simplified design of acoustic carpet cloaks, and it will be shown that these can be built with gratings having only one cavity per unit cell, which greatly simplifies approaches based on metasurfaces which require a large amount of resonators per unit cell. Experiments will be performed to illustrate the functionality of these devices, showing that this simplified approach may open new lines in the design of more advanced devices for the control of acoustic and other waves.

\begin{figure}[ht!]
\begin{center}
\includegraphics[width=\linewidth]{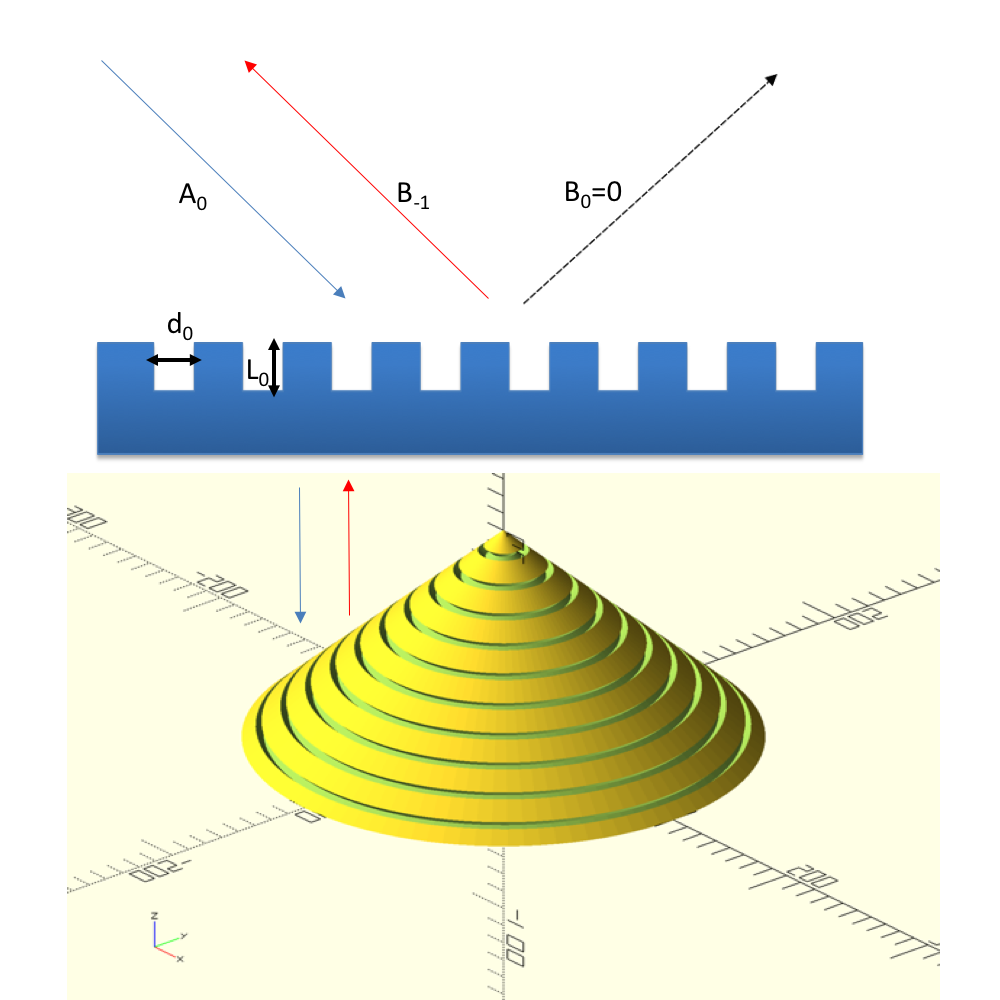}%
\caption{\label{fig:schematics} Schematic representation of the structure considered in this work: An acoustic grating (upper panel) is designed in such a way that only the $n=-1$ mode is excited, being the specularly reflected canceled. Giving a conical shape to this grating (lower panel) we can use this grating as a carpet cloak.}
\end{center}
\end{figure}

Figure \ref{fig:schematics} shows a schematic representation of the idea developed in this work. The upper panel shows a diffraction grating operating in a wavelength such that, for a specific incident angle $\theta_0$, we have only two reflected modes, the fundamental one which is specularly reflected and has an amplitude $B_0$, and the $n=-1$ mode, which has an amplitude $B_{-1}$ and it is ``retroreflected''. This condition ca be achieved according to the diffraction condition
\beq
\sin\theta_n=\sin\theta_0\pm \frac{2n\pi}{ka}
\eeq
with $k$ being the wavenumber of the incident plane wave and $a$ the lattice constant of the grating. The angle $\theta_n$ is the angle of the diffracted wave, and in order to be a propagating wave we require that $|\sin\theta_n|\leq1$, which for a given $k$ happens only for a finite number $n=N_P$ of modes. The grating shown in figure \ref{fig:schematics} is designed so that only the $n=0,-1$ modes be propagative, with the additional condition that $\sin\theta_n=-\sin\theta_0$, which defines the operating $ka$ value as
\beq
ka=\frac{\pi}{\sin\theta_0}
\eeq

The grating consists in periodically drilled cavities of length $L_0$ and width $d_0$ in an acoustically rigid surface. It can be shown that with only one cavity per unit cell it is possible to engineer the grating in such a way that the amplitude of the specular reflection $B_0=0$, so that all the reflected energy goes to the diffracted mode $n=-1$ and the grating acts as a perfect retroreflector. The equation that has to be satisfied for the design of such a grating was shown in\cite{torrent2018acoustic} to be
\beq
\cot kL_0= \frac{d_0}{L_0}\sum_{n\neq 0,-1}\frac{k_b}{|q_n|}\text{sinc}^2\left(\frac{|\bm ka+2\pi n\bm x|d_0}{2a}\right)
\eeq
with $\bm k=k_b(\cos\theta_0\bm x+\sin\theta_0\bm z )$ being the incident wavenumber and $q_n=\sqrt{k_b^2-|\bm k+2\pi n/a\bm x|^2}$. 

We can now give the grating the shape illustrated in the lower panel of figure \ref{fig:schematics}, which is a conical surface whose angle is identical to that satisfying the retroreflection condition, so that any wave arriving with a wavevector parallel to the axis of the cone will be retrorreflected in the vertical direction, cancelling in this way the scattering towards the xy plane and, consequently, cloaking the conical object, since the scattering direction will be the same as if the conical object was not there.

\begin{figure}[ht!]
\begin{center}
\includegraphics[width=\linewidth]{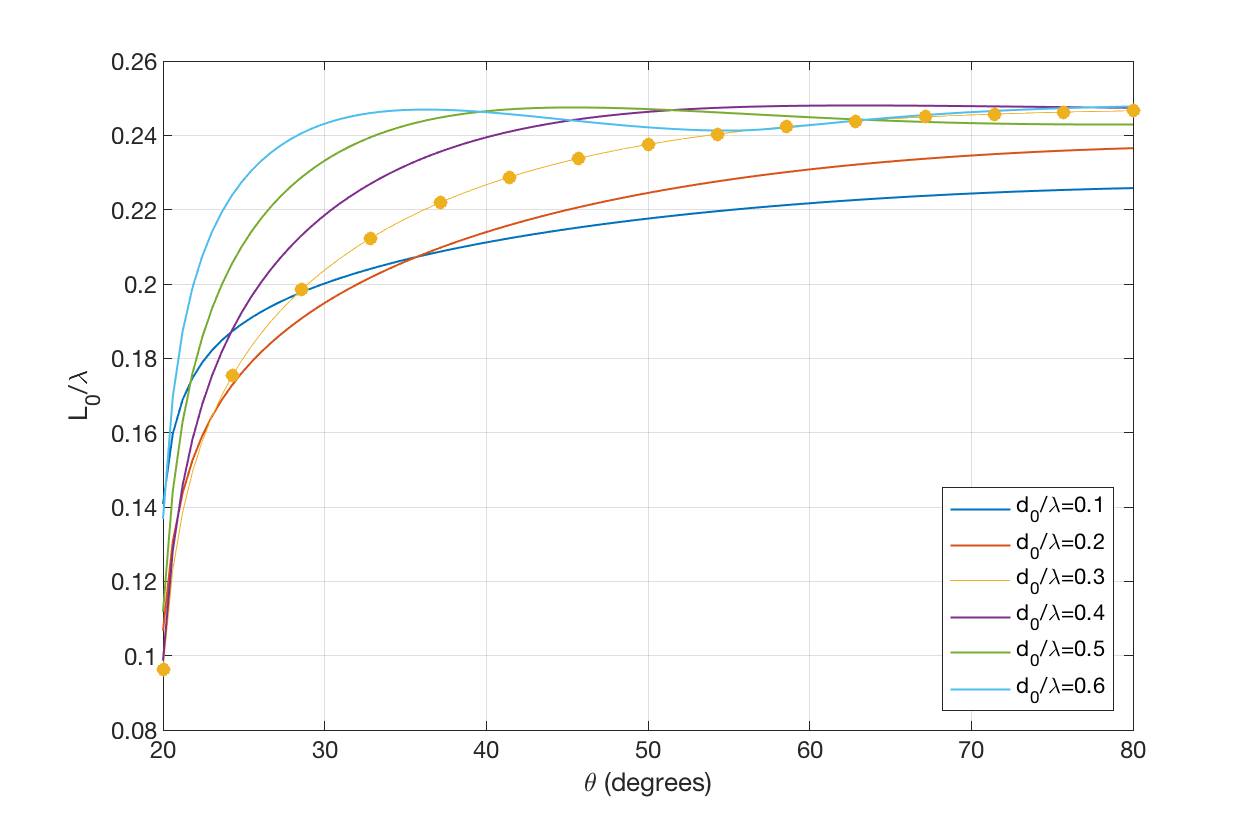}%
\caption{\label{fig:lalpha} Required cavity length $L_0$ as a function of the angle of incidence to cancel the specular reflection of the grating for different values of the cavity width $d_0$.}
\end{center}
\end{figure}

Figure \ref{fig:lalpha} shows the length of the groove $L_0$ relative to the operating wavelength $\lambda$ that we need to satisfy the retroreflection condition as a function of the angle of incidence $\theta_0$. It has to be pointed out that the minimum incidence angle that allows for this condition to be satisfied is given when the $n=1$ mode begins to be propagative, which corresponds to $\sin\theta_0=1/3$ or $\theta_0\approx 20^\circ$. Results are shown for different groove width $d_0$ and the dotted yellow line corresponds to $d_0/\lambda=0.3$, which is the thickness we have employed in our work. It is clear however that the trends of the length $L_0$ are very similar and the variation quite smooth.

\begin{figure}[ht!]
\begin{center}
\includegraphics[width=\linewidth]{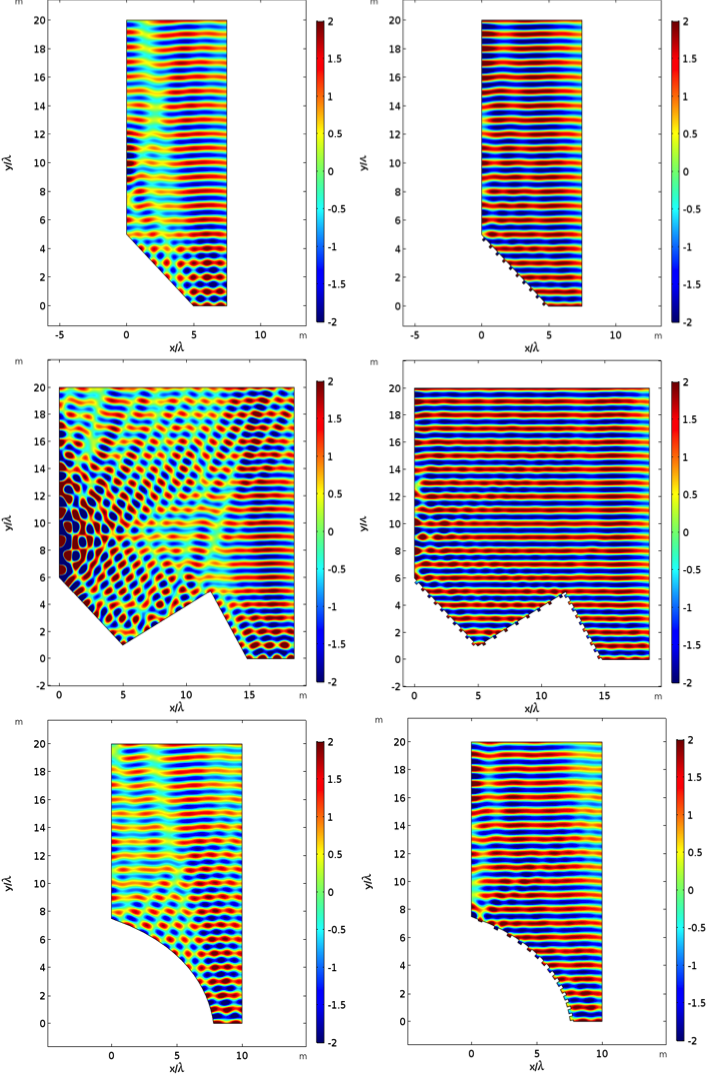}%
\caption{\label{fig:fields} Different carpet cloaks based on engineered diffraction grating. Left panels show the scattering due to the flat object and right panels show the same surface with a drilled grating. We can see a perfect performance of the cloak for a single angle grating (upper panel), a three-angle grating (mid panel) and a multi angle grating (lower panel).}
\end{center}
\end{figure}

Figure \ref{fig:fields} show full wave simulations performed with the commercial software COMSOL of different cloaks designed with this approach, assuming an axisymmetric geometry. A plane wave parallel to the vertical axis (the $z$ axis) arrives from the top of the figure and it is scattered by the objects placed over an acoustically rigid base. The upper panel shows the same conical structure discussed before, with the conical surface making an angle $\theta_0=\pi/4$ with the vertical axis and the propagation direction of the incident field. Left panel shows the scattering of this object with a rigid surface and right panel shows the cloak effect due to the efficient grating design. Clearly the scattering towards the horizontal direction, which corresponds to the specular reflection, has been completely cancelled and only the retroreflection is observed, with an obvious cloaking effect. The mid panel shows that the grating does not need to be uniform to perform properly the cloaking effect, and in this case there are three different angles of the surface which corresponds to three different angles of incidence ($\theta_0=\pi/4,\pi/6$ and $\pi/3$) therefore each surface will have a different length $L_0$ of the grooves. As before, the scattering due to the flat surface (left) is completely cancelled by the grating (right). Finally, the lower panel shows a nearly curved surface in which each angle contains a unit cell, so that each of these surfaces has a different groove length. The retroreflection effect is clear here as well, as can be seen due to the cancelation of the scattering by the surface and the cloak. It is interesting however that, although the design is based on diffraction gratings, which are periodically structured surfaces, the effect still remains even when we use only one unit cell, and we change locally not only the angle of the surface but also the period of the cell. 

\begin{figure}[ht!]
\begin{center}
\includegraphics[width=\linewidth]{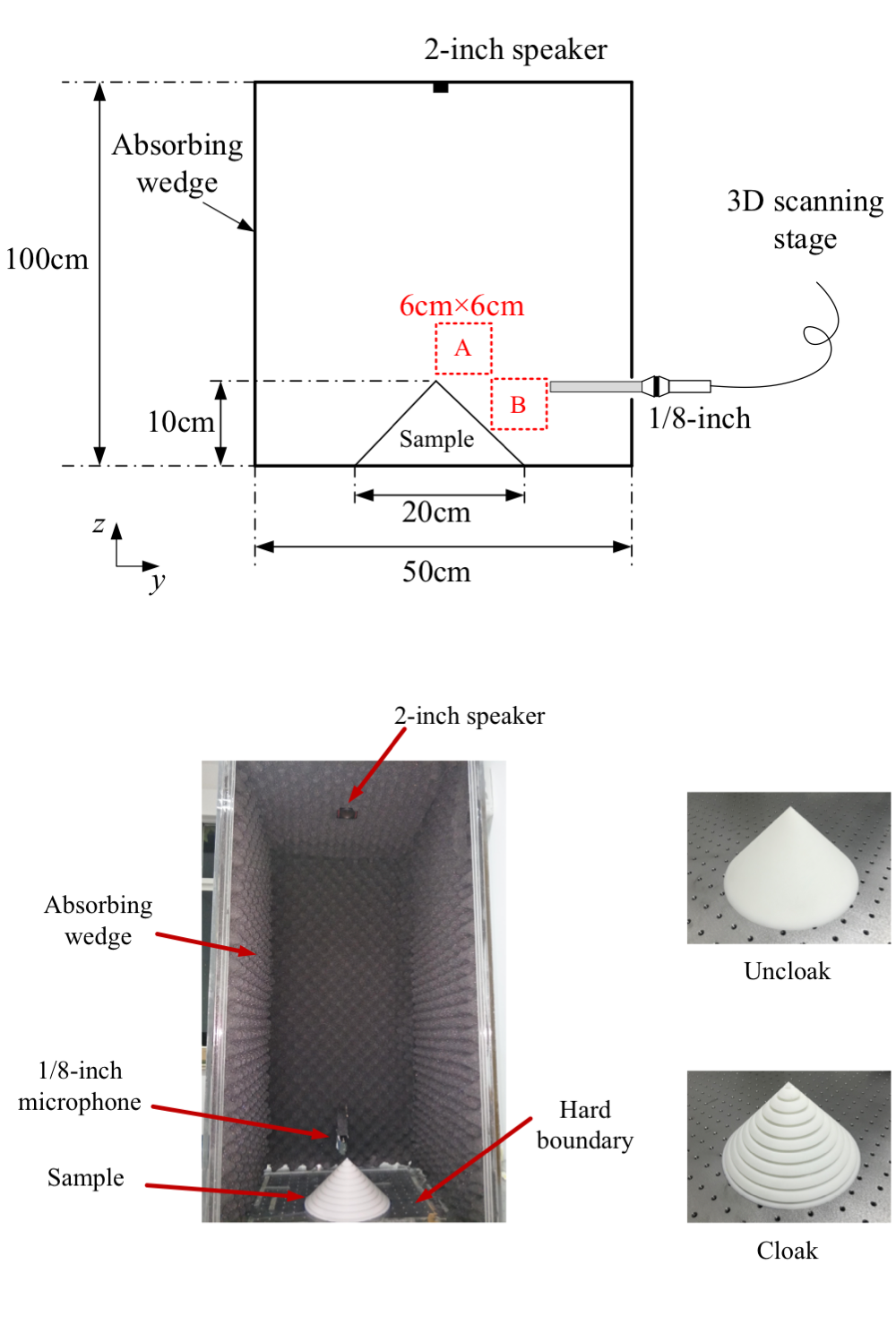}%
\caption{\label{fig:exp} Schematic representation of the experimental set up (upper panel) and pictures of the samples and the chamber (lower panel).}
\end{center}
\end{figure}

The cloaking effect was tested experimentally by means of the first example of figure \ref{fig:fields}. We fabricated samples by using 3D-printing technology, via laser sintering stereo-lithography (SLA) and with photosensitive resin, where the manufacturing precision is 0.1 mm. The wave fields mapping measurements were performed in a man-made anechoic room ($0.5m\times0.5m \times 1 m$) with absorbing wedges (thickness 5 cm) installed on the walls. A continuous sound wave with a center frequency of 17 kHz and 6.4 kHz span was emitted from a two-inch loudspeaker located on the top of the anechoic room. A 1/8-inch Br\"uel \& Kj\ae r  microphone (type 2670) was connected to a three-dimensional moving stage for recording the pressure fields. The measured signal from the microphone and the source signal were connected to the Br\"uel \& Kj\ae r LAN-XI Data Acquisition Hardware (type 3160-A-042) for obtaining the amplitude and phase of the mapping pressure fields. In order to reveal the invisibility effect of the acoustic cloak, we measured two rectangular regions (A and B as shown in Fig. \ref{fig:exp}) on the top and right of samples, respectively. The measured regions ($6\times 6 cm^2$) were meshed into 900  squares with a spacing of 2 mm.

\begin{figure}[ht!]
\begin{center}
\includegraphics[width=\linewidth]{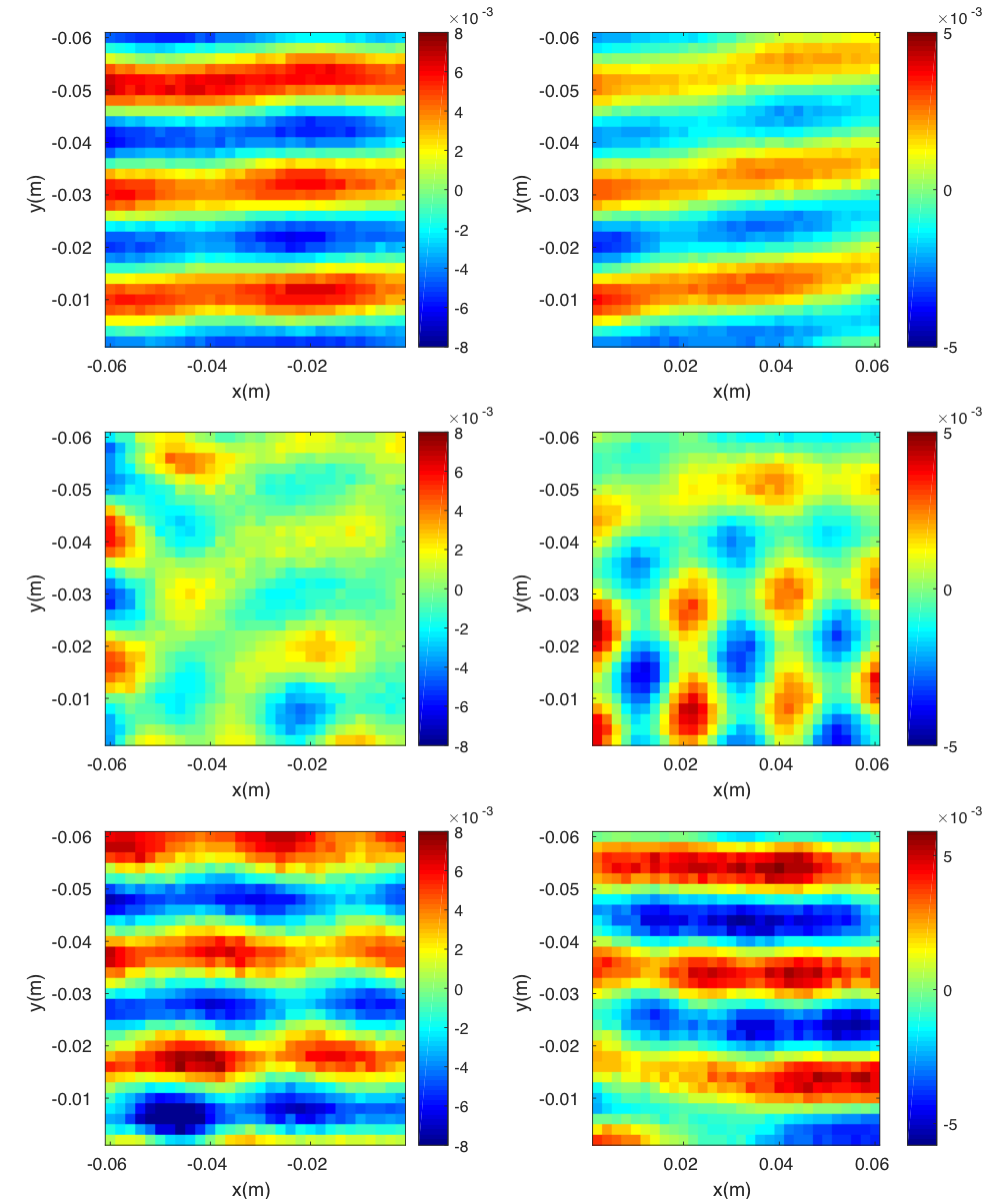}%
\caption{\label{fig:fields_exp} Experimental measurements of the real part of the acoustic field in regions A (left panels) and B (right panels) showed in figure \ref{fig:exp}. We can see how a plane wave is excited and measured without object (upper panels),  this wave is scattered when we put the flat conical surface showed in figure \ref{fig:exp} (mid panels) and finally when the conical surface is structured by the grating the scattered field almost disappears (lower panels).}
\end{center}
\end{figure}

Figure \ref{fig:fields_exp} shows the experimentally measured wave forms for the scanning regions A (left panels) and B (right panels). We can see how when there is no sample in the chamber (upper panel) a nearly plane wave propagates in free space. Once we put the uncloaked object (mid panel) this free field is strongly scattered, and a clear interference pattern can be seen, specially in region B. When the uncloaked object is replaced by the cloak, we see how the free field is nearly recovered (lower panels), cancelling in this way the scattered field towards the undesired direction. It is remarkable the simplicity of this device in comparison with others with similar functionalities, which allow us to build a relatively large sample with strong scattering cross section, as can be seen from both the simulations and the experiments showed in Figs. \ref{fig:fields} and \ref{fig:fields_exp}. 

In summary, we have demonstrated that engineered diffraction gratings can be used to efficiently design carpet cloaks, with the remarkable property of being simpler structures than those presented so far in the literature. It has been shown as well that the grating properties still remains even when the geometry of the grating is modified like giving it a conical shape or changing locally the period of the grating, which allowed us to design three dimensional axisymmetric cloaks with different shapes. The conical cloak was experimentally tested and an excellent functionality was observed. The results found in this work can be easily exported not only to other mechanical waves but also in the electromagnetic domain, where cloaks and similar devices based on complex metasurfaces are currently being investigated.

\acknowledgments
 D.T. acknowledges financial support through the ``Ram\'on y Cajal'' fellowship and by the U.S. Office of Naval Research under Grant No. N00014-17-1-2445.
X.S. and Y.L. acknowledge the support from National Science Foundation of China under Grant No. 11704284 and Shanghai Pujiang Program under Grant No. 17PJ1409000. Y.L. and Y.J. acknowledge the start-up funds from Tongji University. Y.J. and X.F contribute equally to this work.

\end{document}